\documentclass[seceq]{ptptex}

\usepackage{graphicx}

\title{\hfill {\Large TPJU-03/2007}
\\~~~
\\
Remarks on $N_c$ dependence of decays of exotic baryons%
}

\author{
Karolina \textsc{Pie{\'s}ciuk}$^{~}$\footnote{ e-mail address:
yessien@gmail.com}
and
Micha{\l} \textsc{Prasza{\l}owicz}$^{~}$\footnote{ e-mail address:
michal@if.uj.edu.pl}
}

\inst{
$~$ M. Smoluchowski Institute of Physics,
Jagellonian University, Reymonta 4, 30-049 Krak{\'o}w, Poland\\
}

\abst{%
We calculate the $N_c$ dependence of the decay widths of exotic eikosiheptaplet
within the framework of Chral Quark Soliton Model. We also discuss generalizations
of regular baryon representations for arbitrary $N_c$.
}

\begin{document}

\maketitle

\section{Introduction}

One of the most puzzling results of the chiral quark-soliton model ($\chi$%
QSM) for exotic baryons consists in a very small hadronic decay width \cite{Diakonov:1997mm},
governed by the decay constant $G_{\overline{10}}$. While the small mass of
exotic states is rather generic for all chiral models
\cite{Diakonov:1997mm,BieDo,prasz}
the smallness of the decay width appears as a subtle cancelation of three
different terms that contribute to $G_{\overline{10}}$. Decay width in solitonic models
\cite{ANW} 
is calculated in terms of a matrix element $\mathcal{M}$ of the collective
axial current operator corresponding to the emission of a pseudoscalar meson $%
\varphi$ \cite{Diakonov:1997mm}
-- see  Ref. \citenum{Weigel:2007yx}
for criticism of this approach:%
\begin{equation}
\hat{O}_{\varphi}^{(8)}=3\sum_{i=1}^{3}\left( G_{0}\,D_{\varphi
i}^{(8)}-G_{1}\,d_{ibc}\,D_{\varphi b}^{(8)}\hat{S}_{c}-\frac{G_{2}}{\sqrt{3}%
}\,D_{\varphi8}^{(8)}\hat{S}_{i}\right) \times p_{\varphi}^{i}.
\label{Ophi}
\end{equation}
For notation see Ref. \citenum{Diakonov:1997mm}. Constants $G_{0,1,2}$ are constructed
from the so called \emph{%
moments of inertia} that are calculable in $\chi$QSM. The decay width is given as
\begin{equation}
\mathit{\Gamma}_{B\rightarrow B^{\prime}+\varphi}=\frac{1}{8\pi}\frac{%
p_{\varphi}}{M\,M^{\prime}}\overline{\mathcal{M}^{2}}=\frac{1}{8\pi }\frac{%
p_{\varphi}^{3}}{M\,M^{\prime}}\overline{\mathcal{A}^{2}}.   \label{gammadef}
\end{equation}
The \textquotedblleft bar\textquotedblright\ over the amplitude squared
denotes averaging over initial and summing over final spin (and, if
explicitly indicated, over isospin).

For $B^{(\overline{10})}\rightarrow B^{\prime(8)}+\varphi$ for spin "up" and
$\vec{p}_{\varphi}=(0,0,p_{\varphi})$ we have%
\renewcommand{\arraystretch}{0.7}
\begin{equation}
\mathcal{M}=\left\langle 8_{1/2},B^{\prime}\right\vert \hat{O}%
_{\varphi}^{(8)}\left\vert \overline{10}_{1/2},B\right\rangle =-\frac{3G_{%
\overline{10}}}{\sqrt{15}}\left(
\begin{array}{cc}
8 & 8 \\
\varphi & B^{\prime}%
\end{array}
\right\vert \left.
\begin{array}{c}
\overline{10} \\
B%
\end{array}
\right) \times p_{\varphi}   \label{Mdef}
\end{equation}
\renewcommand{\arraystretch}{1.0}
and
\begin{equation}
G_{\overline{10}}=G_{0}-G_{1}-\frac{1}{2}G_{2}.   \label{G10bar}
\end{equation}
In order to have an estimate of the width (\ref{gammadef}) the authors of
Ref. \citenum{Diakonov:1997mm} calculated $G_{\overline{10}}$
in the nonrelativistic limit \cite{limit} of $%
\chi$QSM and got $G_{\overline{10}}\equiv0$. It has been shown that this
cancelation between terms that scale differently with $N_{c}$ ($G_{0}\sim
N_{c}^{3/2},\,G_{1,2}\sim N_{c}^{1/2}$) is in fact consistent with large $%
N_{c}$ counting \cite{Praszalowicz:2003tc}, since
\begin{equation}
G_{\overline{10}}=G_{0}-\frac{N_{c}+1}{4}G_{1}-\frac{1}{2}G_{2}
\label{G10barNc}
\end{equation}
where the $N_{c}$ dependence comes from the SU(3) Clebsch-Gordan coefficients
calculated for large $N_{c}$.
In the nonrelativistic limit (NRL):%
\begin{equation}
G_{0}=-(N_{c}+2)\,G,\quad G_{1}=-4G,\quad G_{2}=-2G,\quad G\sim N_{c}^{1/2}.
\end{equation}

In this paper we ask whether the similar cancelation takes place for the
decays of
$27$ of spin $1/2$ and $3/2$.
We also discuss the possible modifications of the $N_{c}$ dependence of the
decay width due to the different choice of the large $N_{c}$ generalizations
of regular SU(3) multiplets.

\section{Baryons in large $N_{c}$ limit}

Soliton is usually quantized as quantum mechanical
symmetric top with two moments of inertia $I_{1,2}$:%
\begin{equation}
M_{B}^{(\mathcal{R)}}=M_{\text{cl}}+\frac{1}{2I_{1}}S(S+1)+\frac{1}{2I_{2}}%
\left( C_{2}(\mathcal{R})-S(S+1)-\frac{N_{{c}}^{2}}{12}\right) +\delta
_{B}^{(\mathcal{R)}}.
\end{equation}
Here $S$ denotes baryon spin, $C_{2}(\mathcal{R})$ the Casimir operator for
the SU(3) representation $\mathcal{R}=(p,q)$:
\begin{equation}
C_{2}(\mathcal{R})=\frac{1}{3}\left( p^{2}+q^{2}+pq+3(p+q)\right)
\label{Cas}
\end{equation}
and quantities $\delta_{B}^{(\mathcal{R})}$ denote matrix elements of the
SU(3) breaking hamiltonian:
\begin{equation}
\hat{H}^{\prime}=\frac{N_{c}}{3}\sigma+\alpha D_{88}^{(8)})+\beta Y+\frac{%
\gamma}{\sqrt{3}}D_{8A}^{(8)}\hat{J}_{A}.   \label{Hsplit}
\end{equation}
Model parameters that can be found in Ref. \citenum{Blotz:1992pw}%
\[
\alpha=-\frac{N_{c}}{3}(\sigma+\beta),\quad\beta=-m_{s}{\frac{K_{2}}{I_{2}}}%
,\;\gamma=2m_{s}\left( {\frac{K_{1}}{I_{1}}}-{\frac{K_{2}}{I_{2}}}\right)
,\;\sigma={\frac{2}{N_{c}}}{\frac{m_{s}}{m_{u}+m_{d}}}\Sigma_{\pi N}
\]
scale with $N_{c}$ in the following way:%
\begin{equation}
i_{1,2}={3I_{1,2}}/N_{c} \quad\text{where}\quad i_{1,2}\sim \mathcal{O}%
(N_{c}^{0}),\;\sigma,\beta,\gamma\sim\mathcal{O}(m_{s}N_{c}^{0}).
\label{Ncdep}
\end{equation}
Here $\Sigma_{\pi N}$ is pion-nucleon sigma term and $m_{q}$ denote current
quark masses. Numerically $\sigma>\left\vert \beta\right\vert ,\left\vert
\gamma\right\vert $.

\begin{figure}[h]
\begin{center}
\includegraphics[scale=1.1]{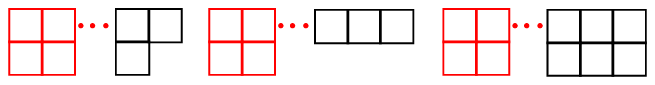}
\end{center}
\caption{{\protect\footnotesize {Adding $\overline{3}$ diquarks to regular
SU(3) baryon representations $8$, $10$ and $\overline{10}$ corresponds to
the representation set of Fig.1.}}}
\label{fig:young1}
\end{figure}

So far we have specified \emph{explicit }$N_{c}$ dependence (\ref{Ncdep})
that follows from the fact that model parameters are given
in terms of the quark loop. Another type of the $N_{c}$ dependence comes
from the constraint \cite{SU3SM} that selects SU(3)$_{\text{flavor}}$
representations $\mathcal{R}=(p,q)$ containing states with hypercharge $%
Y_{R}=N_{c}/3$. Therefore for arbitrary $N_{c}$
ordinary baryon representations have to be extended and one has to specify
which states correspond to the physical ones. Usual choice \cite{largereps}%
\begin{equation}
"8"=\left( 1,(N_{c}-1)/2 \right) ,\;"10"=\left( 3,(N_{c}-3)/2%
\right) ,\;"\overline{10}"=\left( 0,(N_{c}+3)/2 \right) ,
\label{choice1}
\end{equation}
depicted in Fig.~\ref%
{fig:choice1} corresponds -- in the quark language -- to the case when
each time when $N_{c}$ is increased by 2, a spin-isospin singlet (but
charged) $\overline{3}$ diquark is added,
as depicted in Fig.~\ref{fig:young1}.

\begin{figure}[h]
\begin{center}
\includegraphics[scale=1.4]{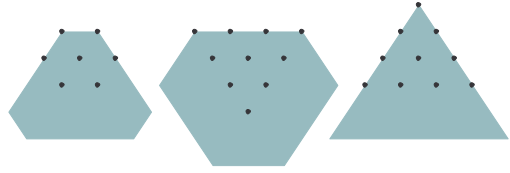}
\end{center}
\caption{{\protect\footnotesize {Standard generalization of SU(3) flavor
baryon representations for arbitrary $N_{c}$}}}
\label{fig:choice1}
\end{figure}

\vspace{0.2cm}

Extension (\ref{choice1}) leads to (\ref{G10barNc}). It
implies that mass differences between centers of
multiplets scale differently with $N_{c}$:%
\begin{equation}
\Delta_{10-8}=\frac{3}{2I_{1}}\sim\mathcal{O}(1/N_{c}),\quad\Delta _{%
\overline{10}-8}=\frac{N_{c}+3}{4I_{2}}\sim\mathcal{O}(1).   \label{split1}
\end{equation}
The fact that $\Delta_{\overline{10}-8}\ne 0$ in large $N_{c}$
limit triggered recently discussion on the validity of the
semiclassical quantization for exotic states \cite{Pobylitsa:2003ju}.
Since in the chiral limit the momentum $p_{\varphi}$ of the outgoing meson 
scales according to (%
\ref{split1}), overall $N_{c}$ dependence of the decay width is strongly
affected by its third power (\ref{gammadef}):
\begin{equation}
\mathit{\Gamma}_{B\rightarrow B^{\prime}+\varphi}\sim\frac{1}{N_{c}^{2}}%
\mathcal{O}(\overline{\mathcal{A}^{2}})\mathcal{O(}p_{\varphi}^{3}).
\label{gamNc}
\end{equation}

Phenomenologically, however, scaling (\ref{split1}) is not sustained.
Indeed, meson momenta in $\Delta$ and $\Theta$ decays are almost identical
(assuming $M_{\Theta}^{(\overline{10})}\simeq1540$ MeV):%
\begin{equation}
p_{\pi}\simeq225\;\text{MeV},\;p_{K}\simeq268\;\text{MeV}.   \label{momexp}
\end{equation}

Unfortunately, going off SU(3)$_{\text{flavor}}$ limit does not help.
Explicitly:
\begin{align}
\delta^{(8)} & =\frac{N_{c}}{3}\sigma+\frac{(N_{c}-3)}{3}\beta+\frac {%
(N_{c}-2)\alpha+\frac{3}{2}\gamma}{N_{c}+7}+\left( \beta+\frac {%
3(N_{c}+2)\alpha-\frac{1}{2}(2N_{c}+9)\gamma}{(N_{c}+3)(N_{c}+7)}\right) Y
\notag \\
& +\frac{\left( 6\alpha+(N_{c}+6)\gamma\right) }{(N_{c}+3)(N_{c}+7)}\left(
\frac{Y^{2}}{4}-I(I+1)\right) =3\sigma+2\beta-\sigma Y+\ldots \\
\delta^{(10)} & =\frac{N_{c}}{3}\sigma+\frac{(N_{c}-3)(N_{c}+4)}{%
(N_{c}+1)(N_{c}+9)}\alpha+\frac{N_{c}-3}{3}\beta+\frac{5(N_{c}-3)}{%
2(N_{c}+1)(N_{c}+9)}\gamma  \notag \\
& +\left( \beta+\frac{3(N_{c}-1)\alpha-\frac{5}{2}(N_{c}+3)\gamma}{%
(N_{c}+1)(N_{c}+9)}\right) Y=3\sigma+2\beta-\sigma Y+\ldots \\
\delta^{(\overline{10})} & =\frac{N_{c}}{3}\sigma+\frac{N_{c}(N_{c}-3)}{%
(N_{c}+3)(N_{c}+9)}\alpha+\frac{N_{c}-3}{3}\beta-\frac{3(N_{c}-3)}{%
2(N_{c}+3)(N_{c}+9)}\gamma  \notag \\
& +\left( \beta+\frac{6N_{c}\alpha-9\gamma}{2(N_{c}+3)(N_{c}+9)}\right)
Y=5\sigma+4\beta-\sigma Y+\ldots
\end{align}
where $\ldots$ denote terms $\mathcal{O}(1/N_{c})$, $Y$ and $I$ denote
{\em physical} hypercharge and isospin.

Interestingly in all cases in the large $N_{c}$ limt, $m_{s}$ splittings are
proportional to the hypercharge differences only. In this limit $\Sigma
-\Lambda$ splitting in the octet is zero and this degeneracy is lifted in
the next order at $\mathcal{O}(1/N_{c})$. This explains the
smallness of $\Sigma-\Lambda$ mass difference. Additionally $%
\delta_{N}^{(8)}\simeq\delta_{\Delta}^{(10)}$ up to higher order terms $%
\mathcal{O}(1/N_{c}^{2})$, however $\delta_{\Theta}^{(\overline{10}%
)}-\delta_{N}^{(8)}$ $\simeq\sigma+2\beta>0$. This implies that

\begin{align}
M_{\Theta}^{(\overline{10})}-M_{N}^{(8)} & =\frac{3}{2I_{2}}-\frac{1}{20}%
\alpha+\beta-\frac{3}{40}\gamma\rightarrow\frac{3}{4i_{2}}+\sigma +2\beta+%
\mathcal{O}(1/N_{c}),\quad  \notag \\
M_{\Delta}^{(10)}-M_{N}^{(8)} & =\frac{3}{2I_{1}}-\frac{7}{40}\alpha -\frac{%
21}{80}\gamma\rightarrow\mathcal{O}(1/N_{c}).   \label{mass-diffs}
\end{align}
The first equation shows that the $\Theta-N \ne 0$  in
the large $N_{c}$ limit even if $m_{s}$ corrections are included. We will
come back to this problem in the last section.

\section{Decay constants of twentysevenplet for large $N_{c}$}

In this section we shall consider decays of eikosiheptaplet (27-plet)
\begin{equation}
"27"=\left( 2,(N_{c}+1)/2 \right)
\end{equation}
that can have either spin $1/2$ or $3/2$, the latter being lighter. Mass
differences read%
\begin{align}
\Delta_{27_{3/2}-8} & =\frac{3}{2I_{1}}+\frac{N_{c}+1}{4I_{2}}\sim\mathcal{O}%
(1),\;\,\Delta_{27_{1/2}-8}=\frac{N_{c}+7}{4I_{2}}\sim\mathcal{O}(1),  \notag
\\
\Delta_{27_{3/2}-10} & =\frac{N_{c}+1}{4I_{2}}\sim\mathcal{O}(1),\;\qquad
\quad\Delta_{27_{1/2}-10}=-\frac{3}{2I_{1}}+\frac{N_{c}+7}{4I_{2}}\sim%
\mathcal{O}(1),  \notag \\
\Delta_{27_{3/2}-\overline{10}} & =\frac{3}{2I_{1}}-\frac{1}{2I_{2}}\sim%
\mathcal{O}(1/N_{c}),\;\Delta_{27_{1/2}-\overline{10}}=\frac{1}{I_{2}}\sim%
\mathcal{O}(1/N_{c}).
\end{align}
Matrix elements for the decays of eikosiheptaplet (with $S_{3}=1/2$) read:%
\begin{align}
\mathcal{A}(B_{27_{3/2}} & \rightarrow B_{8}^{\prime}+\varphi)=3\left(
\begin{array}{cc}
8 & "8" \\
\varphi & B^{\prime}%
\end{array}
\right\vert \left.
\begin{array}{c}
"27" \\
B%
\end{array}
\right) \sqrt{\frac{8(N_{c}+5)}{9(N_{c}+3)(N_{c}+9)}}\times G_{27},  \notag
\\
\mathcal{A}(B_{27_{3/2}} & \rightarrow B_{10}^{\prime}+\varphi)=-3\left(
\begin{array}{cc}
8 & "10" \\
\varphi & B^{\prime}%
\end{array}
\right\vert \left.
\begin{array}{c}
"27" \\
B%
\end{array}
\right) \sqrt{\frac{(N_{c}-1)(N_{c}+7)}{9(N_{c}+1)(N_{c}+3)(N_{c}+9)}}\times
F_{27},  \notag \\
\mathcal{A}(B_{27_{3/2}} & \rightarrow B_{\overline{10}}^{\prime}+\varphi)=3%
\left(
\begin{array}{cc}
8 & "\overline{10}" \\
\varphi & B^{\prime}%
\end{array}
\right\vert \left.
\begin{array}{c}
"27" \\
B%
\end{array}
\right) \sqrt{\frac{2(N_{c}+1)(N_{c}+7)}{3(N_{c}+3)(N_{c}+9)}}\times E_{27},
\end{align}
and
\renewcommand{\arraystretch}{1.3}
\begin{equation*}
\begin{array}{|c|rl|c|}
\hline
\text{Decay} & \text{Large }N_{c} & \text{NRL} &
\begin{array}{c}
\text{Scaling} \\
\text{in NRL}%
\end{array}
\\ \hline
27_{3/2}\rightarrow 8_{1/2} & G_{27}=G_{0}-\frac{N_{c}-1}{4}G_{1} & =-3G &
N_{c}^{1/2} \\
27_{3/2}\rightarrow 10_{3/2} & F_{27}=G_{0}-\frac{N_{c}-1}{4}G_{1}-\frac{3}{2%
}G_{2} & =0 & 0 \\
27_{3/2}\rightarrow \overline{10}_{1/2} & E_{27}=G_{0}+G_{1} & =-(N_{c}+6)G
& N_{c}^{3/2} \\ \hline
\end{array}%
\end{equation*}%
\renewcommand{\arraystretch}{1} For $S=1/2$ and $S_{3}=1/2$ we have:%
\begin{align}
\mathcal{A}(B_{27_{1/2}}& \rightarrow B_{8}^{\prime }+\varphi )=-3\left(
\begin{array}{cc}
8 & "8" \\
\varphi  & B^{\prime }%
\end{array}%
\right\vert \left.
\begin{array}{c}
"27" \\
B%
\end{array}%
\right) \sqrt{\frac{(N_{c}+1)(N_{c}+5)}{9(N_{c}+3)(N_{c}+7)(N_{c}+9)}}\times
H_{27},  \notag \\
\mathcal{A}(B_{27_{1/2}}& \rightarrow B_{10}^{\prime }+\varphi )=-3\left(
\begin{array}{cc}
8 & "10" \\
\varphi  & B^{\prime }%
\end{array}%
\right\vert \left.
\begin{array}{c}
"27" \\
B%
\end{array}%
\right) \sqrt{\frac{8(N_{c}-1)}{9(N_{c}+3)(N_{c}+9)}}\times G_{27}^{\prime },
\notag \\
\mathcal{A}(B_{27_{1/2}}& \rightarrow B_{\overline{10}}^{\prime }+\varphi
)=3\left(
\begin{array}{cc}
8 & "\overline{10}" \\
\varphi  & B^{\prime }%
\end{array}%
\right\vert \left.
\begin{array}{c}
"27" \\
B%
\end{array}%
\right) \frac{N_{c}+4}{\sqrt{9(N_{c}+3)(N_{c}+9)}}\times H_{27}^{\prime },
\end{align}
\renewcommand{\arraystretch}{1.5}
\begin{equation*}
\begin{array}{|c|rl|c|}
\hline
\text{Decay} & \text{Large }N_{c} & \text{NRL} &
\begin{array}{c}
\text{Scaling} \\
\text{in NRL}%
\end{array}
\\ \hline
27_{1/2}\rightarrow8_{1/2} & {H_{27}=G}_{0}{-}\frac{N_{c}+5}{4}{G_{1}+\frac {%
3}{2}G_{2}} & {=0} & 0 \\
27_{1/2}\rightarrow10_{3/2} & G_{27}^{\prime}={G}_{0}{-\frac{N_{c}+5}{4}G_{1}%
} & {=3G} & N_{c}^{1/2} \\
27_{1/2}\rightarrow\overline{10}_{1/2} & H_{27}^{\prime}=G_{0}+\frac{2N_{c}+5%
}{2N_{c}+8}G_{1}+\frac{3}{2N_{c}+8}G_{2} & =-\frac{(N_{c}+3)(N_{c}+7)}{%
N_{c}+4}G & N_{c}^{3/2} \\ \hline
\end{array}%
\end{equation*}
\renewcommand{\arraystretch}{1}

In order to calculate the $N_{c}$ behavior of the width we have to know the $%
N_{c}$ dependence of the flavor Clebsch-Gordan coefficients that
depend on the states involved. For the decays into 8 and 10 the
only possible channels are $\Theta_{27}\rightarrow N(\Delta)+K$, and the
pertinent Clebsches do not depend on $N_{c}$. For the decays into $\overline{%
10}$ we have $\Theta_{27}\rightarrow\Theta_{\overline{10}}+\pi$ that scales
like $\mathcal{O}(1)$ and $\Theta_{27}\rightarrow N_{\overline {10}}+K $ that
scales like $\mathcal{O}(1/\sqrt{N_{c}})$. The resulting scaling of $\mathit{%
\Gamma}_{\Theta_{27}\rightarrow B^{\prime}+\varphi}$ calculated from Eq.(\ref%
{gamNc}) reads as follows:

\renewcommand{\arraystretch}{1.3}
\begin{equation*}
\begin{array}{|lcc||lcc|}
\hline
\text{decay of }
& \multicolumn{2}{c||}{N_{c}\text{ scaling}} & \text{decay of } &
\multicolumn{2}{c|}{N_{c}\text{ scaling}} \\
\;\;\Theta_{27_{3/2}}
& \text{exact} & \text{NRL} & \;\;\Theta_{27_{1/2}} & \text{exact} & \text{NRL} \\
\hline
\rightarrow N_{8}+K & \mathcal{O}(1) & \mathcal{O}%
(1/N_{c}^{2}) & \rightarrow N_{8}+K & \mathcal{O}(1) & 0 \\
\rightarrow\Delta_{10}+K & \mathcal{O}(1) & 0 &
\rightarrow\Delta_{10}+K & \mathcal{O}(1) & \mathcal{O}%
(1/N_{c}^{2}) \\
\rightarrow N_{\overline{10}}+K & \mathcal{O}(1/N_{c}^{3})
& \mathcal{O}(1/N_{c}^{3}) & \rightarrow N_{\overline{10}}+K
& \mathcal{O}(1/N_{c}^{3}) & \mathcal{O}(1/N_{c}^{3}) \\
\rightarrow\Theta_{\overline{10}}+\pi & \mathcal{O}%
(1/N_{c}^{2}) & \mathcal{O}(1/N_{c}^{2}) & \rightarrow
\Theta_{\overline{10}}+\pi & \mathcal{O}(1/N_{c}^{2}) & \mathcal{O}%
(1/N_{c}^{2}) \\ \hline
\end{array}%
\end{equation*}
\renewcommand{\arraystretch}{1} Interestingly, we see that whenever the
exact scaling is $\mathcal{O}(1)$, the nonrelativistic cancelation (exact or
partial) lowers the power of $N_{c}$, whereas in the case when the width has
\emph{good} behavior for large $N_{c}$, there is no NRL cancelation.

\section{Alternative choices for large $N_{c}$ multiplets}

So far we have only considered the "standard" generalization (\ref{choice1})
of baryonic SU(3)$_{\text{flavor}}$ representations for large $N_{c}$. This
choice is based on the requirement that generalized baryonic states have
physical spin, isospin and strangeness, however their hypercharge and charge
are not physical \cite{largereps}. Moreover, the generalization of the octet
is not
selfadjoint and antidecuplet is not complex conjugate of decuplet. Some
years ago it has been proposed to consider alternative schemes \cite{dul}.

\begin{figure}[h]
\begin{center}
\includegraphics[scale=1.4]{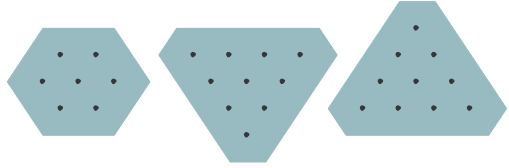}
\end{center}
\caption{{\protect\footnotesize {Generalization of SU(3) flavor
representations in which octet is selfadjoint}}}
\label{fig:choice2}
\end{figure}

\vspace{0.4cm}
\begin{figure}[h]
\begin{center}
\includegraphics[scale=1.1]{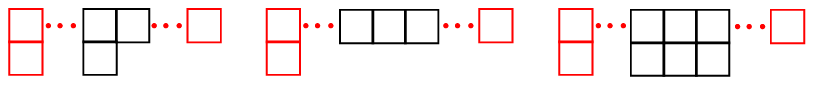}
\end{center}
\caption{{\protect\footnotesize {Adding triquarks to regular SU(3) baryon
representations $8$, $10$ and $\overline{10}$ corresponds to the
representation set of Fig.3.}}}
\label{fig:young2}
\end{figure}

If we require the generalized octet to be self-adjoint we are led to the
following set of representations%
\begin{equation}
"8"=\left( N_{c}/{3},N_{c}/{3}\right) ,\quad "10"=\left(
(N_{c}+6)/{3},(N_{c}-3)/{3}\right) ,\quad "\overline{10}"="10"^{\ast}
\label{choice2}
\end{equation}%
that are depicted in Figs. \ref{fig:choice2} and \ref{fig:young2}.
This means that we enlarge $N_{c}$ in steps of $3$ adding each
time a $uds$ triquark. Generalized states have physical isospin, hypercharge (and
charge), but unphysical strangeness and spin that is of the order of $N_{c}$%
. With this choice both $\Delta _{10-8}$, $\Delta
_{\overline{10}-8} \ne 0$
in large $N_{c}$ limit:
\begin{equation}
\Delta _{10-8}=\left( N_{c}/{6}-1\right)/{I_{1}},\quad \Delta
_{\overline{10}-8}=\left( N_{c}/{6}-1\right)/{I_{2}}.
\label{split2}
\end{equation}%
With this power counting we can calculate large $N_{c}$ approximation of the
meson momenta in the decays of $\Delta $ and $\Theta $:%
\begin{align}
\Delta & \rightarrow N\qquad p_{\pi }=\sqrt{(M_{\Delta }-M_{N})^{2}-m_{\pi
}^{2}}=256\;\text{MeV,}  \notag \\
\Theta & \rightarrow N\qquad p_{K}=\sqrt{(M_{\Theta }-M_{N})^{2}-m_{K}^{2}}%
=339\;\text{MeV}  \label{momch2}
\end{align}%
that are much closer to the physical values (\ref{momexp}) than (\ref{split1}%
).
\begin{figure}[h]
\begin{center}
\includegraphics[scale=1.4]{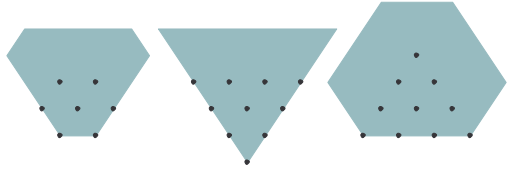}
\end{center}
\caption{{\protect\footnotesize {Generalization of SU(3) flavor
representations in which decuplet is fully symmetric $(0,q)$.}}}
\label{fig:choice3}
\end{figure}
\vspace{0.1cm}
\begin{figure}[h]
\begin{center}
\includegraphics[scale=1.1]{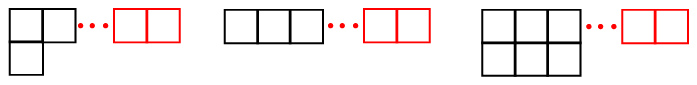}
\end{center}
\caption{{\protect\footnotesize {Adding sextet diquarks to regular SU(3)
baryon representations $8$, $10$ and $\overline{10}$ corresponds to the
representation set of Fig.5.}}}
\label{fig:young3}
\end{figure}

Finally let us mention a third possibility in which we require generalized
decuplet to be a completely symmetric SU(3)$_{\text{flavor}}$ representation
for arbitrary $N_{c}$. This leads to (see Figs. \ref{fig:choice3} and \ref%
{fig:young3}):%
\begin{equation}
"8"=\left( N_{c}-2,1\right) \qquad"10"=\left( N_{c},0\right) \qquad"%
\overline{10}"=\left( N_{c}-3,3\right) .  \label{choice3}
\end{equation}
Interestingly this choice has a smooth limit to the one flavor case. In the
quark language it amounts to adding a symmetric diquark to the original SU(3)%
$_{\text{flavor}}$ representation when increasing $N_{c}$ in steps of $2$.
As seen from Fig. \ref{fig:choice3} physical states are situated at the
bottom of infinite
representations (\ref{choice3}) and therefore have unphysical strangeness,
charge (hypercharge) and also spin.

The mass splittings for this choice read%
\begin{equation}
\Delta _{10-8}=N_{c}/\;{2I_{1}},\quad \Delta _{\overline{10}-8}=
3/\;{2I_{2}}.  \label{split3}
\end{equation}%
Here the generalized decuplet remains split from the $"8"$, while $\Delta _{%
\overline{10}-8}\rightarrow 0$ for large $N_{c}$. The phase space factor for
$\Theta $ decay is therefore suppressed with respect to the one of $\Delta $.

\section{Summary}

In this short note we have shown that very small width of exotic baryons
-- if they exist -- cannot be explained by the standard $N_{c}$ counting
alone. Certain degree of \emph{nonrelativisticity} is needed to ensure
cancelations between different terms in the decay constants. This phenomenon
observed firstly for antidecuplet, is also operative for the decays of
eikosiheptaplet. We have shown that in $\chi $QSM in the nonrelativistic
limit all decays are suppressed for large $N_{c}$. Exact cancelations occur
for $\Theta _{27_{3/2}}\rightarrow \Delta _{10}+K$ and $\Theta
_{27_{1/2}}\rightarrow N_{8}+K$, leading $N_{c}$ terms cancel for $\Theta
_{27_{3/2}}\rightarrow N_{8}+K$ and $\Theta _{27_{1/2}}\rightarrow \Delta
_{10}+K$. For $27\rightarrow \overline{10}$ there are no cancelations, but
the phase space is $N_{c}^{-3}$ suppressed.

We have also briefly discussed nonstandard generalizations of regular baryon
representations for arbitrary $N_{c}$. For $N_{c}>3$ bayons are no longer
composed from 3 quarks and therefore they form large SU(3)$_{\text{flavor}}$
representations that reduce to octet, decuplet and antidecuplet for $N_{c}=3$%
. The standard way to generalize regular baryon representations is to add
antisymmetric antitriplet diqaurk when $N_{c}$ is increased in intervals of
2. This choice fulfils many reasonable requirements; most importantly for
SU(2)$_{\text{flavor}}$ these representations form regular isospin
multiplets. However, representations (\ref{choice1}) do not obey conjugation
relations characteristic for regular representations. Therefore we have
proposed generalization (\ref{choice2}) that satisfies conjugation
relations. Most important drawback of (\ref{choice2}) is that spin $S\sim
N_{c}$ that contradicts semiclassical quantization. Nevertheless as a result
meson momenta emitted in $10$ and $\overline{10}$ decays scale in the same
way with $N_{c}$ (\ref{momch2}), consistently with "experimental" values (%
\ref{momexp}), whereas for (\ref{choice1}) the scaling is different (\ref%
{split1}).

\bigskip

\noindent{\bf Acknowledgements}

One of us (MP) is grateful to the organizers of the Yukawa International
Symposium (YKIS2006) for hospitality during this very successful
workshop.

%


\end{document}